\documentclass{article}
\usepackage{spconf,amsmath,graphicx}
\usepackage{amssymb}
\usepackage{multirow}
\usepackage{xcolor}

\title{Enhancing Expressivity Transfer in Textless Speech-to-Speech Translation}
\name{Jarod Duret$^1$, Benjamin O'Brien$^1$, Yannick Est{\`e}ve$^1$, Titouan Parcollet$^2$}
\address{
  $^1$LIA - Avignon Universite, France\\
  $^2$University of Cambridge, United-Kingdom}

\copyrightnotice{979-8-3503-0689-7/23/\$31.00~\copyright2023 IEEE}
\begin{document}
%
\maketitle
\begin{abstract}
Textless speech-to-speech translation systems are rapidly advancing, thanks to the integration of self-supervised learning techniques.
However, existing state-of-the-art systems fall short when it comes to capturing and transferring expressivity accurately across different languages.
Expressivity plays a vital role in conveying emotions, nuances, and cultural subtleties, thereby enhancing communication across diverse languages.
To address this issue this study presents a novel method that operates at the discrete speech unit level and leverages multilingual emotion embeddings to capture language-agnostic information. Specifically, we demonstrate how these embeddings can be used to effectively predict the pitch and duration of speech units in the target language. Through objective and subjective experiments conducted on a French-to-English translation task, our findings highlight the superior expressivity transfer achieved by our approach compared to current state-of-the-art systems.

\end{abstract}
\begin{keywords}
speech translation, prosody prediction, speech generation
\end{keywords}
\section{Introduction}
\label{sec:intro}
In today's interconnected world, speech-to-speech translation (S2ST) technology can help bridge the communication gap between people speaking different languages by enabling effective communication across diverse languages and cultures. Nevertheless, existing speech-to-speech translation systems frequently fall short of retaining the subtleties of expressiveness embedded within the speaker's original message. Developing speech-to-speech translation systems capable of capturing the emotional and expressive dimensions of spoken language is crucial to improve the naturalness of speech generation. Conventional speech-to-speech translation systems rely on cascaded approaches~\cite{lavie1997, nakamura2006} that follow a two-step approach first converting the source speech into a textual representation in the target language domain. This can be accomplished by using automatic speech recognition (ASR) followed by machine translation (MT), or by using an end-to-end speech-to-text translation (S2T) system~\cite{berard2016, Jia2019}. The resulting text output is then transformed into a speech using text-to-speech (TTS). 

More recently, a textless direct speech-to-speech translation (S2ST) approach has been proposed which relies on discrete speech units~\cite{Lee2022direct}. 
This approach is particularly valuable when translating from an unwritten language and/or to an unwritten language.
Furthermore, it has been observed that this method is also highly effective for languages that possess a written form~\cite{Lee2022direct,Lee2022textless}.
This technique is designed to effectively capture the linguistic content of the target speech while minimizing the impact of the speaker's prosodic features.
Previous study~\cite{Polyak2021} has shown that the utilization of discrete speech units successfully disentangles linguistic content from the influence of prosodic characteristics and speaker identity.


Another challenge in preserving expressivity in speech-to-speech translation is the lack of parallel annotated speech data. The recent approach introduced in~\cite{Lee2022textless} is designed to address this lack of paired speech data by focusing on the linguistic context. However, this approach does not address the issues of preserving emotions and other non-linguistic information contained in the source language speech.

Inspired by a recent work on prosody reconstruction from multilingual speech representation~\cite{Duret2023}, we propose an approach that aims to build a speech-to-speech translation system that preserves the expressivity without the need for parallel speech data. This approach consists of training a multilingual emotion embedding extractor used to compute an emotion embedding from an utterance and exploit it for speech resynthesis. In our work, we extend this approach to address emotion preservation in textless speech-to-speech translation. We compute an emotion embedding from the source utterance and use it to condition the duration and pitch predictor models used for generating the target utterance from a discrete speech unit representation.




\section{Related Work}
\label{sec:related}

\textbf{Spoken Language Modeling from audio}.
Generative spoken language modeling from audio is a task that involves acquiring the acoustic and linguistic characteristics of a language solely from raw audio data, without any accompanying text or labeled information. 
In~\cite{Lakhotia2021}, the authors proposed to leverage advancements in self-supervised speech representation learning to discover discrete speech units and subsequently use them in downstream tasks. They demonstrate that speech generation can be achieved by sampling sequences of these discovered units from a unit-discovery model and synthesizing them into a coherent speech waveform using a unit-to-speech model. Building upon this work, in~\cite{Polyak2021}, the authors demonstrated the effectiveness of utilizing self-supervised learned discrete speech units for generating high-quality speech. Furthermore, a comparable approach and speech representation scheme were employed for the purpose of textless speech emotion conversion through translation~\cite{Kreuk2022} and for prosody reconstruction using a multilingual speech representation~\cite{Duret2023}. In this study, we leverage a similar approach and speech representation scheme to encode target speech in order to train a speech-to-unit translation model.
\newline

\noindent
\textbf{Speech-to-speech translation}.
Most speech-to-speech translation systems rely on cascaded approaches that require intermediate text representation. 
This makes them unusable for languages without written forms or datasets containing only speech alignments. Recent research on S2ST is new, exploring scenarios involving speech-to-speech translation (S2ST) that does not rely on intermediate text representation. In~\cite{Jia2019}, an attention-based sequence-to-sequence neural network was proposed to enable direct speech translation without the need for intermediate text representation. The model was trained end-to-end, mapping speech spectrograms from a source language to target spectrograms in another language.
Additionally, the authors introduced a variation that aimed to transfer the voice characteristics of the source speaker to the translated speech. However, as the model was trained on synthetic data, the voice transfer capabilities did not achieve comparable results to those observed in a similar text-to-speech context. In subsequent work, ~\cite{Lee2022direct} introduces a direct S2ST system based on self-supervised discrete representations. The proposed approach exhibits enhanced performance compared to its predecessor, unfortunately, it remains constrained by the utilization of synthetic data. Furthermore, it is important to note that this study did not emphasize the exploration of paralinguistic information. More recently, ~\cite{Lee2022textless} tackles direct S2ST by following~\cite{Lee2022direct} and focuses on training the system with real-world data on multiple language pairs.
Previous studies in the field of direct speech-to-speech translation have predominantly concentrated on improving the quality of the translation, disregarding the paralinguistic dimension and expressivity transfer.
In contrast, the current study aims to build a speech-to-speech translation framework that can transfer the expressivity from one language into another.

\section{ARCHITECTURE}
\begin{figure*}[ht!]
\centering
\includegraphics[width=0.9\textwidth]{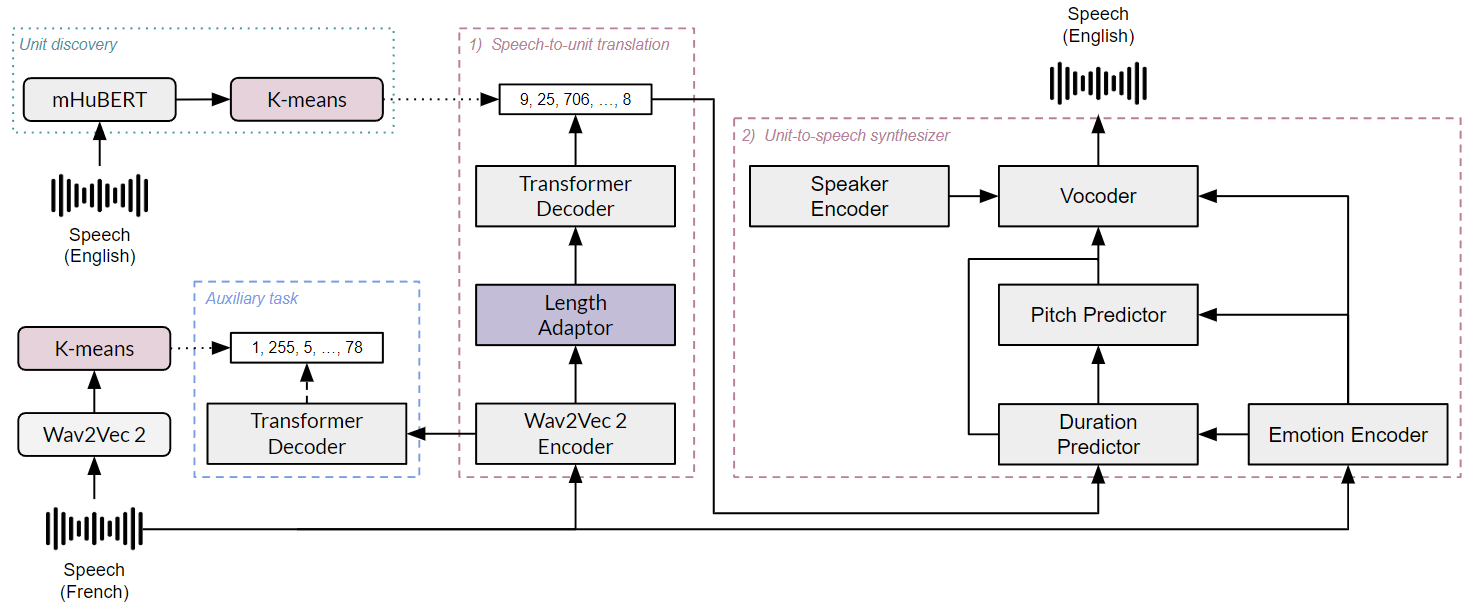}
\caption{Illustration of our proposed speech-to-speech translation model. First, the input speech is translated into a sequence of discrete units by the speech-to-unit translation model (1). Next, we predict duration and F0 before feeding them to a unit-to-speech model (2). Duration Predictor, Pitch Predictor, and the unit-to-speech model are conditioned by the emotion embedding extracted from the source speech by the emotion encoder. The speaker is encoded using a 1-hot vector directly in the unit-to-speech model.}
\label{fig:architecture}
\end{figure*}

Our speech-to-speech translation framework does not require parallel speech data for speaker and expressivity modeling, enables the translation of speech while maintaining the inherent expressive content, and can generate speech in the target language with multiple voices.
The proposed framework can be decomposed into two parts.
First, a speech-to-unit translation model (Section \ref{subsec:translation}), composed of a speech encoder and an acoustic decoder. Secondly, a unit-to-speech synthesizer (Section \ref{subsec:unit_2_unit}), composed of an emotion encoder, a speaker encoder, a duration predictor, a pitch predictor and a speech vocoder.
The following subsections describe each component of the proposed S2ST framework while the overall architecture is illustrated in Figure~\ref{fig:architecture}.



\subsection{Speech-to-unit translation model}
\label{subsec:translation}
The following describes the speech-to-unit translation (S2UT) model~(1) depicted in Figure~\ref{fig:architecture}.
In order to capture the linguistic content, particularly pseudo-phonetic information present in speech, we employ a pre-trained self-supervised learning (SSL) model to extract raw speech features from the audio signal, namely multilingual HuBERT (mHuBERT) \cite{Lee2022textless} for English and  Wav2Vec 2.0 \cite{Conneau2020} for French.
Wav2Vec 2.0 and mHuBERT models are pre-trained in a self-supervised manner and produce continuous representations for every 20-ms frame. To extract the sequence of speech units, a k-means clustering is applied to the raw speech features and the learned $K$ cluster centroids are used to transform audio into a sequence of cluster indices at every 20ms of the input audio signal. For English speech, we extracted representations from the $11^{th}$ layer of mHuBERT model and set \begin{math} k = 1000 \end{math} as used in \cite{Lee2022textless} for speech-to-speech translation. 
For French speech, we extracted representations from the  $11^{th}$ layer of Wav2Vec2-XLSR model and set \begin{math} K = 1000\end{math}. Following \cite{Lee2022textless}\cite{Kreuk2022}, as a way of speeding up training and inference time, we experimented with reducing a sequence of units to a sequence of unique units by removing consecutive duplicated units (e.g., 0, 0, 1, 1, 1, 2 → 0, 1, 2). We denote such sequences as “reduced”.

We build the S2UT model by adapting the transformer encoder-decoder framework presented in~\cite{Popuri2022}.
As an encoder, we chose a large Wav2Vec 2.0 pre-trained on 7.6K hours of French speech (1.8K Males / 1.0K Females / 4.8K unknown)~\footnote{https://huggingface.co/LeBenchmark/wav2vec2-FR-7K-large}. In contrast to the encoder, the decoder consists of $6$ transformer layers with a random initialization for each transformer decoder weight. To alleviate the mismatch between the length of the source speech and the reduced target units, we introduced an adaptor layer of a single 1-D convolutional layer with stride 2 between the encoder and the decoder. We combined the Wav2Vec 2.0 encoder, the adaptor along with the transformer decoder and we finetune the whole model end-to-end. Following\cite{Lee2022textless}, we explored an auto-encoding style auxiliary task by adding a separate transformer decoder as auxiliary task to help the model converge during training. This separate transformer consists of $3$ transformer layers, which are trained to predict the discrete units sequence of the source speech as the target.

\subsection{Unit-to-speech model}\label{subsec:unit_2_unit}
The following section describes all components of the unit-to-speech (U2S) model~(2) depicted in Figure~\ref{fig:architecture}. \newline

\noindent\textbf{Emotion Encoder}
To capture emotion representations, we use a dedicated encoder specifically trained for emotion recognition tasks within a multilingual context. Our proposed architecture is inspired by~\cite{Duret2023} and involves a pre-trained Wav2Vec2-XLSR encoder, a bottleneck layer, and a dense layer. We fine-tuned both the CNN and Transformer modules of the Wav2Vec2-XLSR model following the approach described in \cite{Wang2021}. The information across the entire source speech sequence is encoded into a single fixed-length vector representation of size $96$ by passing the audio through the bottleneck layer and applying temporal pooling.
\newline

\noindent\textbf{Speaker Encoder}
In order to synthesize speech using the voices of several speakers, we introduce a speaker representation that serves as an additional conditioning factor. Inspired by~\cite{Duret2023}, we optimize the parameters of a fixed-size look-up table.
Although using speaker representations from a pre-trained speaker encoder enables generalization to new and unseen speakers, it is worth noting that such embedding captures a broader range of speaker-related information and it may be slightly less efficient in capturing solely the speaker identity, resulting in a degradation of synthesized speech.
\newline

\noindent\textbf{Duration Predictor}
As the speech-to-unit translation model reduced sequences, we were required to predict the duration of each discrete unit before feeding them to the pitch predictor and unit-to-speech model. For this purpose, we take inspiration from work on TTS~\cite{Ren2020}, where a CNN is used to predict the duration of each phoneme from a phoneme sequence.
Following~\cite{Kreuk2022, Duret2023}, we replaced the phoneme sequence with the reduced discrete unit sequence and predict the number of repetitions for each unit, in order to reconstruct the original sequence. During training, we conditioned the model using an emotion embedding extracted from ground-truth speech and the ground-truth discrete unit duration is used as supervision. At inference time, the emotion embedding is extracted from the speech in the source language.
\newline

\noindent\textbf{Pitch Predictor}
Pitch is an important characteristic of speech prosody, however, due to the non-monotonic alignment characteristic of speech-to-speech translation, a direct extraction of pitch from the source signal is not viable. Thus, an alternative method was required to accurately estimate pitch in this context. To overcome this limitation, we introduce a F0 estimation model to predict the pitch directly from a sequence of speech units. During the training phase, we use the ground-truth speech in order to extract the speech units sequence, and, during inference, we use the output of the S2UT model. Our pitch predictor model is a CNN followed by a linear layer projecting the output to $\mathbb{R}^d$. We apply a sigmoid on the model prediction to output a vector in $[0,1]^d$. During the training phase, the target F0 is extracted using the YAAPT \cite{Kasi2002} algorithm.
Following~\cite{Kreuk2022, Duret2023}, we discretize ranges of F0 values into $d$ bins, represented by one-hot encodings.
Then, we compute the weighted-average of the activated bins in order to expand the output range during the conversion of bins back to F0 values. We apply a normalization on the F0 values using the mean and standard deviation for each speaker.

Like the F0 estimation model, the duration predictor model is conditioned using an emotion embedding extracted from ground-truth speech. The same embedding is used to condition both models.
\newline

\noindent\textbf{Speech synthesis}
Following~\cite{Polyak2021}, we use the HiFi-GAN neural vocoder \cite{Kong2020} to synthesize speech.
HiFiGAN is a generative adversarial network (GAN) that consists of one generator and a set of discriminators.
The generator is a fully convolutional neural network.
Inspired by~\cite{Kreuk2022}, we adapted the generator architecture to take as input a sequence of discrete-unit inflated using the predicted durations, predicted F0, emotion-embedding, and a speaker-embedding. Before feeding the above features into the model, we concatenate them along the temporal axis. The sample rates of unit sequence and F0 are matched by means of linear interpolation, while the speaker-embedding and emotion-label are replicated along the temporal axis.

Regarding the set of discriminators, the model is composed of two modules: a Multi-Scale Discriminators (MSD) and a Multi-Period Discriminators (MPD). The first type operates on different sizes of sliding windows over the input signal, while the latter samples the signal at different periods.

\section{Experimental Setup}

We use the SpeechMatrix~\cite{duquenne2022} corpora for training and evaluating our speech-to-unit translation (S2UT) model. SpeechMatrix consists of $126$ language pairs with a total of $418$ thousand hours of speech from European Parliament recordings. In this study, only French-to-English language pairs were considered, yielding a $1,507$ hours train set.


In addition to the mined speech-to-speech data for training purposes, we extend our evaluation by leveraging labeled public speech datasets obtained from two distinct corpora that cover various domains.
First, Europarl-ST (EPST)~\cite{iranzo2020}, a multilingual corpus containing paired audio-text samples built from recordings of debates from the European Parliament, containing $72$ translation directions in $9$ languages, including French to English direction.
The second dataset is FLEURS~\cite{conneau2023Fleurs}.
Derived from from FLoRes~\cite{goyal2022flores}, FLEURS is an extension that introduces speech recordings for these translated texts, resulting in a collection of speech-to-speech data comprising French to English direction.
FLEURS texts are from English Wikipedia.
During training, we extract a validation set from SpeechMatrix of about $1000$ sample which are not in the test set.
FLEURS validation set is derived from its validation samples.
To compute evaluation scores, we consider only the source speech and target texts, the complete evaluation pipeline is described in section~\ref{subsec:eval}.

The unit-to-speech (U2S) model is separately trained from S2UT model. To train the U2S system for English language, we combine the LJSpeech dataset \cite{Keith2017} and the ESD\cite{Zhou2021} dataset. 
The LJSpeech dataset contains 13,100 short audio clips of a single speaker reading passages from $7$ non-fiction books, with a total duration of approximately $24$ hours. 
ESD is a multilingual emotional database, consisting of 350 parallel utterances spoken by 10 native English and 10 native Chinese speakers (10F, 10M).
In this study, we only consider the English part.

\subsection{Baseline}
To assess the effectiveness of our proposed approach, we build a Baseline model which is composed of a speech-to-unit translation (S2UT) module and a unit-to-speech (U2S) module.
We conduct an analysis by systematically excluding the emotion encoder and pitch predictor from the U2S module. This enables us to quantify the impact and measure the benefits of their inclusion in the overall system.

\subsection{TTS}
In addition to the Baseline and our S2ST model, we also incorporated an English text-to-speech (TTS) model~\cite{kim2021conditional} into our subjective evaluation. 
The TTS model was trained on the identical dataset utilized for training the U2S module.
Its inclusion serves to assess the overall quality of the synthesized speech generated by the TTS model in comparison to our proposed system and to evaluate the effectiveness of expressivity transfer achieved by our proposed system in contrast to the TTS model.

\subsection{Evaluation}
\label{subsec:eval}

\begin{figure}[ht!]
\centering
\includegraphics[width=0.45\textwidth]{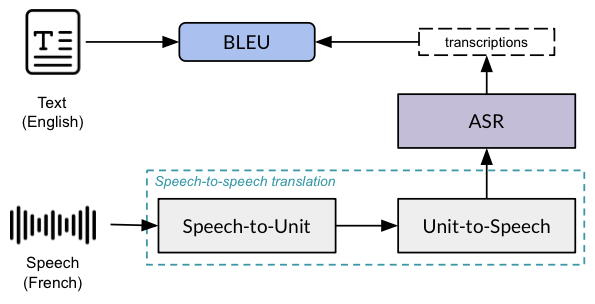}
\caption{An illustration of the evaluation pipeline used for speech-to-speech translation.}
\label{fig:eval}

\end{figure}

Recent work in speech-to-speech translation suggests to evaluate translation quality using the BLEU score.
We start by using an ASR model to compute the transcriptions of the generated speech.
In order to obtain comparable results, we use the same open source ASR model as in~\cite{duquenne2022}.
Then, we compute BLEU score of the ASR decoded text with respect to the reference translations.
We acknowledge that the ASR BLEU score may not be a perfect metric for assessing data quality, as it will be unavoidably influenced by the performance of ASR models.
The complete evaluation pipeline of speech-to-speech translation is illustrated in Figure~\ref{fig:eval}.

In addition to measuring the translation quality via an objective metric, we conduct human listening tests to assess perceptual responses of expressivity transfer from recordings generated by our S2ST model.
We asked 33 people to evaluate two sets of tasks online. A detailed description of the tasks was provided to all evaluators, who had unlimited time to evaluate audio stimuli. Each task was organized similarly, consisting of one pre-trial (excluded from this analysis) and four trials. Each trial contained three synthesized speech recordings produced by Baseline, TTS, and our S2ST framework. 
After listening to each recording, evaluators provided an opinion score on a scale of 1 to 5, where 1 is `Poor' and 5 is `Excellent'. The first task was a Mean Opinion Score (MOS) where evaluators judged the quality of the synthesized speech. The second task was a Multiple Stimuli with Hidden Reference and Anchor (MUSHRA), where evaluators listened to a reference (natural, spoken French) and then judged the expressiveness of the English-translated synthesized speech.

\section{Results}

We first evaluate translation quality of the Baseline and the S2ST model using BLEU score (Section 5.1).
Next, we conduct a subjective evaluation in terms of audio-quality
(MOS) along with expressivity transfer (MUSHRA) and compare the proposed method against the Baseline and TTS (Section 5.2).

\subsection{Speech-to-speech translation}

\begin{table}[tb]
    \caption{BLEU scores on EPST and FLEURS test sets by S2ST models with different settings}
    \label{tab:bleu_scores}
\begin{tabular}{|p{\dimexpr0.30\textwidth-2\tabcolsep-1.3333\arrayrulewidth}|c c|}
\hline
\multirow{2}{*}{Model} & \multicolumn{2}{c|}{BLEU} \\
& EPST & FLEURS  \\
\hline
Synthetic target & 82.6 & 82.7 \\
\hline
    Baseline   & 17.0 & 15.7 \\
    S2ST & 17.3 & 15.9 \\
    Baseline~\emph{multitask} & 16.7 & 14.0 \\
    S2ST~\emph{multitask} & 17.0 & 14.2 \\
\hline
From the literature: & ~ & ~ \\
    SpeechMatrix & 20.7 & 9.8 \\
\hline
\end{tabular}
\end{table}
We investigate the training of a speech-to-speech translation system using both single and multitask learning approaches.
Table~\ref{tab:bleu_scores} summarizes performance of S2ST models on both EPST and FLEURS test sets.
We include the results from SpeechMatrix~\cite{duquenne2022} as references as the exact same ASR models is used for evaluation.
Additionally, we present the BLEU scores calculated for the synthetic target speech to show the impact of ASR errors on the evaluation metric.

First, we compare the proposed S2ST model to the Baseline.
We can see that our S2ST model outperforms the Baseline by 0.3 BLEU on EPST and by 0.2 BLEU on FLEURS, indicating that our approach performs similar or slightly better in terms of translation performance.
We also note that SpeechMatrix achieves an improvement of 3.4 BLEU over the proposed S2ST model on EPST, however, on FLEURS our approach outperforms SpeechMatrix by 6.1 BLEU leading to an average improvement of 1.3 BLEU.
The gap of performance on the FLEURS test set can be attributed in part to the fact that we use an encoder pre-trained on 7000 hours of speech coverings multiples domains compared to SpeechMatrix encoder trained only on European Parliament recording.

Secondly, we explore multitask learning by incorporate an auxiliary task to the Baseline and S2ST model.
In our experimental setup, we observe a decline in performance for both the Baseline and the S2ST model when employing multitask learning.
Specifically, the S2ST model yields a performance of (17 vs. 17.3) on EPST and (14.2 vs. 15.9) on FLEURS.
This suggests that our encoder does not provide significant benefits to the auxiliary task.
Nonetheless, our approach still outperforms the Baseline system for both setups, indicating the effectiveness of our proposed approach.

\subsection{Subjective Evaluation}

Separate linear mixed effects models were used to evaluate MOS and MUSHRA task responses. Using the R-package $lme4$, opinion responses were entered as response variables. Synthesized speech system (3-levels) and speaker sex (2-levels) were entered as fixed factors and participant was entered as a random factor. Chi-squared ($\chi^2_{d,N}$) tests were used to report $p$-values ($Anova$ from the $car$ R-Package) with $d$ degrees of freedom and $N$ samples, i.e., there were $N$ = 486 responses, $d$ = 2 speech systems, and $d$ = 1 speaker sexes. Main effects were reported for task, response, and their interactions with speaker. Estimated marginal means ($emmeans$) were used to conduct pairwise comparisons, where $X \pm Y$ represent mean and standard error, respectively.

\begin{figure*}[h!]
\centering
\includegraphics[width=0.58\textwidth]{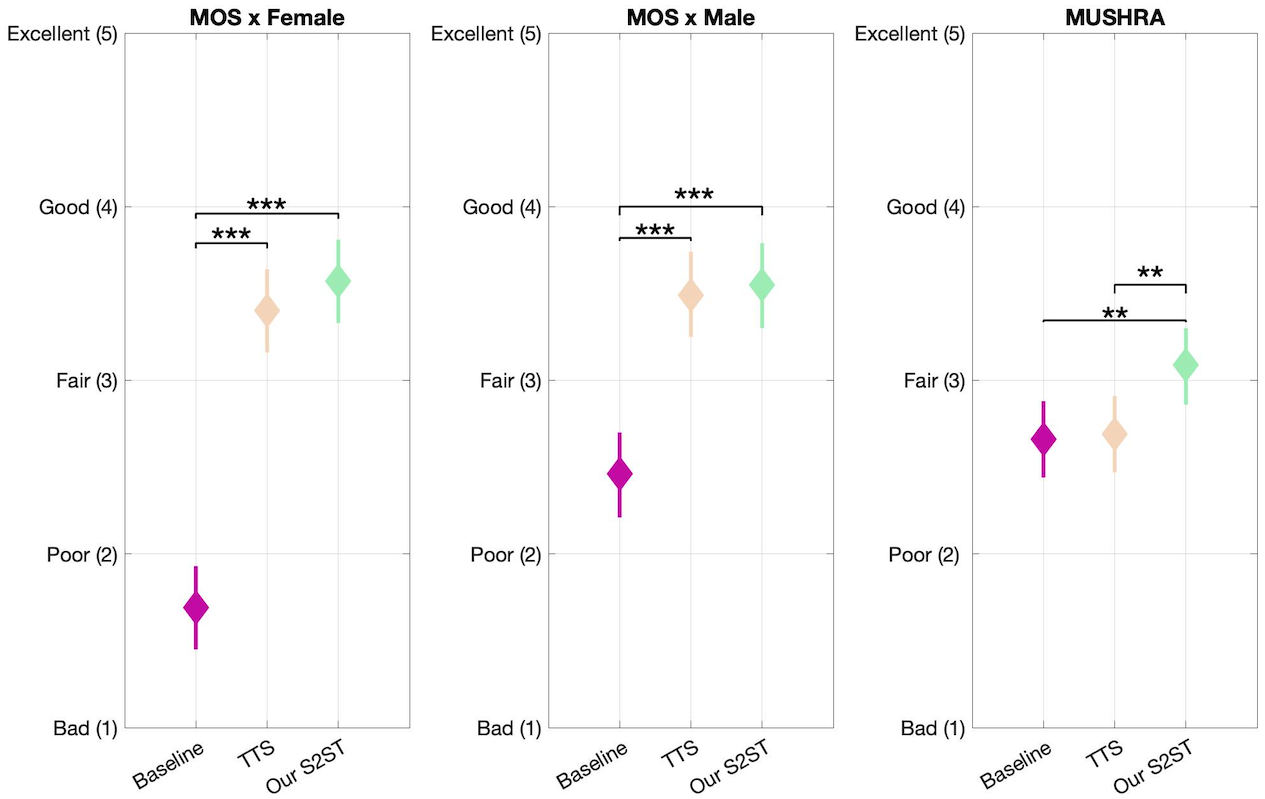}
\caption{MOS (Left-Middle) and MUSHRA (Right) task results. Diamonds and vertical lines represent mean and critical intervals. \{**, ***\} represent $p <$ \{0.01, 0.001\}.}
\label{fig:sub_eval}
\end{figure*}

The results of the MOS task revealed significant main effects on system $\chi^2_{2, 486}$ = 284.17 and speaker sex $\chi^2_{1, 486}$ = 11.25, as well as their interaction $\chi^2_{2, 486}$ = 18.66, $p <$ 0.001. Pairwise comparisons showed that the quality of recordings generated by the Baseline system (2.07 $\pm$ 0.1) had significantly lower opinion scores in comparison to those generated by TTS (3.45 $\pm$ 0.1) and our S2ST model systems (3.56 $\pm$ 0.1), $p <$ 0.001. In comparison to female speech recordings (2.89 $\pm$ 0.09), male speech recordings (3.17 $\pm$ 0.09) had significantly increased scores, $p <$ 0.001, however, these effects were localized to the Baseline system (Figure \ref{fig:sub_eval} Left-Middle).

MUSHRA task results showed a significant main effect on system $\chi^2_{2, 453}$ = 14.27, $p <$ 0.001, but not on speaker sex, $p >$ 0.05. Pairwise comparisons showed that the expressiveness of recordings generated by our S2ST model (3.08 $\pm$ 0.11) had significantly increased higher opinion scores in comparison to those generated by Baseline (2.66 $\pm$ 0.11) and TTS systems (2.69 $\pm$ 0.11), $p <$ 0.01 (Fig. \ref{fig:sub_eval}-Right).

To better understand the MUSHRA task results, OpenSmile was used to extract 88-acoustic features (eGeMAPS \cite{Eyben2010}) that were entered in a forward SLDA with Wilks' Lambda criterion (R function $greedy.wl$ in the $klaR$ R-package). In order to identify which acoustic features distinguished our S2ST model from Baseline and TTS systems, a forward SLDA method was preferred, as it starts from the null hypothesis and incrementally adds new variables with the highest discriminant power based on Wilks' Lambda value until $p >$ 0.01. Based on the six acoustic features selected\footnote{The following acoustic features were selected (with Wilks' lambda and $F$-stat values): slopeUV500{\textunderscore}1500{\textunderscore}sma3nz{{\textunderscore}}amean ($\lambda$: 0.22; $F$: 80.26), slopeV0{\textunderscore}500{\textunderscore}sma3nz{\textunderscore}amean ($\lambda$: 0.06; $F$: 56.2), F0semitoneFrom27{\textunderscore}5Hz{\textunderscore}sma3nz{\textunderscore}stddevRisingSlope ($\lambda$: 0.03; $F$: 17.16), F1bandwidth{\textunderscore}sma3nz{\textunderscore}amean ($\lambda$: 0.02; $F$: 14.98), logRelF0{\textunderscore}H1{\textunderscore}H2{\textunderscore}sma3nz{\textunderscore}amean ($\lambda$: 0.01; $F$: 17.47), and mfcc4V{\textunderscore}sma3nz{\textunderscore}stddevNorm ($\lambda$: 0.01; $F$: 8.88).} from the SLDA, standardized euclidean distances were computed between the reference (French) and synthesized (English) speech recordings produced by Baseline, TTS, and the proposed S2ST model. One-sided ANOVA results revealed a significant effect of system on euclidean distances between reference and synthesized speech $F_{2, 87}$ = 4.11, $p <$ 0.05. Pairwise comparisons showed our S2ST model (2.75 $\pm$ 0.25) had significantly smaller distances in comparison to the TTS system (3.75 $\pm$ 0.25), however, no differences from the Baseline (3.16 $\pm$ 0.25). Finally, Pearson correlation procedures showed a significant relationship between mean opinion and euclidean distance between the reference and our S2ST model ($\rho$ = -0.39, $p <$ 0.05).

There are several takeaways from our subjective evaluations. First our S2ST framework produced speech recording that were perceived to have higher quality in comparison to those produced by the Baseline system. Next it outperformed both Baseline and TTS systems in terms of producing recordings that conveyed speaker expressivity. The euclidean distances of a select set of acoustic features (6) extracted from reference and our S2ST model speech recordings were found to be significantly smaller in comparison to TTS system and negatively correlated to opinion scores.

\section{Conclusions}
In this paper, we have addressed the crucial challenge of preserving expressivity in speech-to-speech translation systems. Our proposed approach leverages multilingual emotion embeddings, resulting in significant advancements in retaining the nuances of expressiveness during textless translation. The experimental results have demonstrated the superior expressivity transfer achieved by our method compared to state-of-the-art systems, highlighting its effectiveness.

Moreover, our speech-to-speech translation framework has produced speech recordings that were perceived by humans to have higher quality in terms of conveying speaker expressivity, surpassing both our speech-to-speech Baseline and text-to-speech systems. Importantly, we have maintained the translation quality at a level similar to that of state-of-the-art textless speech-to-speech translation systems.

Looking ahead, future research directions involve exploring the incorporation of additional paralinguistic information, optimizing the generation of speech discrete units for this task, and expanding the approach to other language pairs, particularly unwritten ones.

\section{Acknowledgments}
This work received funding from the European SELMA project (grant N°957017).

\bibliographystyle{IEEEbib}
\bibliography{refs}
 
\end{document}